\documentclass[aps, prl, reprint]{revtex4-1}
\usepackage{graphicx}
\usepackage{color}
\usepackage{amssymb}
\usepackage{afterpage}
\usepackage{amsmath}
\usepackage{tabularx}
\usepackage{bm}
\usepackage{hyperref}
\usepackage{ltxtable}
\usepackage{natbib} 
\usepackage[caption=false]{subfig}
\usepackage{longtable}
\usepackage[belowskip=-15pt,aboveskip=0pt]{}
\usepackage{float}
\usepackage{IEEEtrantools}
\newcommand{\be}{\begin{equation}}
\newcommand{\ee}{\end{equation}} 
\newcommand{\bes}{\begin{equation*}}
\newcommand{\ees}{\end{equation*}} 
\newcommand{\bea}{\begin{eqnarray}}
\newcommand{\eea}{\end{eqnarray}}

\usepackage[usenames,dvipsnames]{xcolor}
\hypersetup{
colorlinks,
citecolor=blue,
filecolor=blue,
linkcolor=blue,
urlcolor=blue}

\begin{document}
\title{Restoring Chaos Using Deep Reinforcement Learning}

\author{Sumit Vashishtha}
\email{svashishtha2019@fau.edu}
\affiliation{Department of Ocean and Mechanical engineering, Florida Atlantic University, Boca Raton, FL 33431, USA}

\author{Siddhartha Verma}
\email{vermas@fau.edu}
\affiliation{Department of Ocean and Mechanical engineering, Florida Atlantic University, Boca Raton, FL 33431, USA}

\date{\today}

\begin{abstract}
{A catastrophic bifurcation in non-linear dynamical systems, called crisis, often leads to their convergence to an undesirable non-chaotic state after some initial chaotic transients. Preventing such behavior has proved to be quite challenging. We demonstrate that deep Reinforcement Learning (RL) is able to restore chaos in a transiently-chaotic regime of the Lorenz system of equations. Without requiring any a priori knowledge of the underlying dynamics of the governing equations, the RL agent discovers an effective perturbation strategy for sustaining the chaotic trajectory. We analyze the agent's autonomous control-decisions, and identify and implement a simple control-law that successfully restores chaos in the Lorenz system. Our results demonstrate the utility of using deep RL for controlling the occurrence of catastrophes and extreme-events in non-linear dynamical systems.}
\end{abstract}

\maketitle

Chaos is desirable and advantageous in many situations. For instance, in mechanics, exciting the chaotic motion of several modes spreads energy over a wide frequency range~\cite{georgiou1996slow}, thereby preventing undesirable resonance. Chaotic advection in fluids enhances mixing, as chaos brings about an exponential divergence of fluid packets that are initially in close proximity~\cite{haken1981chaos}. In biology, the absence of chaos may lead to an emergence of synchronous dynamics in the brain, which can result in epileptic seizures~\cite{sackellares2000epilepsy}. Moreover, the absence of chaos may also indicate the presence of other pathological conditions~\cite{goldberger1990chaos, yang1995preserving}.

In some cases, Chaos can become transient in nature, where the dynamics eventually converge to non-chaotic attractors. The typical route by which this happens is known as a crisis~\cite{grebogi1983crises}, where for certain parameter-values of the non-linear system, a chaotic-attractor collides with its basin-boundary and becomes a saddle. A saddle has a fractal structure with infinitely many gaps along its unstable-manifold. Any initial condition attracted towards this chaotic-attractor-turned-saddle escapes to an external periodic- or a fix-point-attractor. Such transient-chaos is often undesirable, and has been conjectured to be the culprit for phenomena such as voltage collapse in electric power systems~\cite{dobson1989towards} and species extinction in ecology~\cite{mccann1994nonlinear}. It also plays a crucial role in governing the dynamics of shear flows in pipes and ducts at low Reynolds numbers~\cite{avila2011onset, kreilos2014increasing}. Given the importance of these phenomena, controlling transient-chaos is a pressing issue.

Some attempts to restore chaos in such scenarios have been made in the past. \citet{yang1995preserving} maintained chaos in transiently chaotic regimes of one- and two-dimensional maps using small perturbations. Their method relied on accurate  analytical knowledge of the dynamical system, and required a priori phase-space knowledge of escape regions from chaos. Another method utilized the natural dynamics around the saddle~\cite{schwartz1996sustaining, dhamala1999controlling}, where small regions near a chaotic-saddle through which trajectories escape were identified. Then a set of ``target" points in these regions were found, which yield trajectories that can stay near the chaotic saddle for a relatively long time. When the solution trajectory falls in this escape region, it is perturbed to the nearest target point so that the trajectory can persist near the chaotic saddle for a long time. The identification of such escape regions and target points can be challenging, and requires either an a priori computation of the probability distribution of escape times in different regions of state-space~\cite{schwartz1996sustaining}, or information from the return map constructed from local maxima or minima of a measured time series~\cite{dhamala1999controlling}. Such approaches become difficult for high-dimensional dynamical systems, and have been illustrated for 2D maps/flows at the most. One particular control technique that worked for the 3D Lorenz-system was described in~\citet{capeans2017partially}. The method was based on finding a certain control-perturbation set in the phase space, called a ``safe set", which avoids the escape of the trajectories to the fix-points. Identifying such a safe set can be prohibitively expensive computationally, and such safe sets may not exist for all dynamical systems. 

In recent years, a machine-learning technique called deep Reinforcement Learning (RL) has shown great promise in control-optimization problems~\cite{silver2016mastering}, and it has been successfully used to uncover complex underlying physics in Navier-Stokes simulations of fish-swimming~\cite{verma2018efficient}. The aim of this letter is to illustrate the utility of deep RL in determining small control-perturbations to parameters of the Lorenz system~\cite{lorenz1995essence}, such that a sustained chaotic behavior is maintained despite the uncontrolled dynamics being transiently-chaotic. In doing so, no prior analytical knowledge about the dynamical system, and no special schemes to find escape regions, target points and safe sets will be employed. The RL algorithm is able to autonomously determine an optimal strategy to restore chaos, by continually interacting with the dynamical-system.

\begin{figure}
\includegraphics[width=0.7\linewidth]{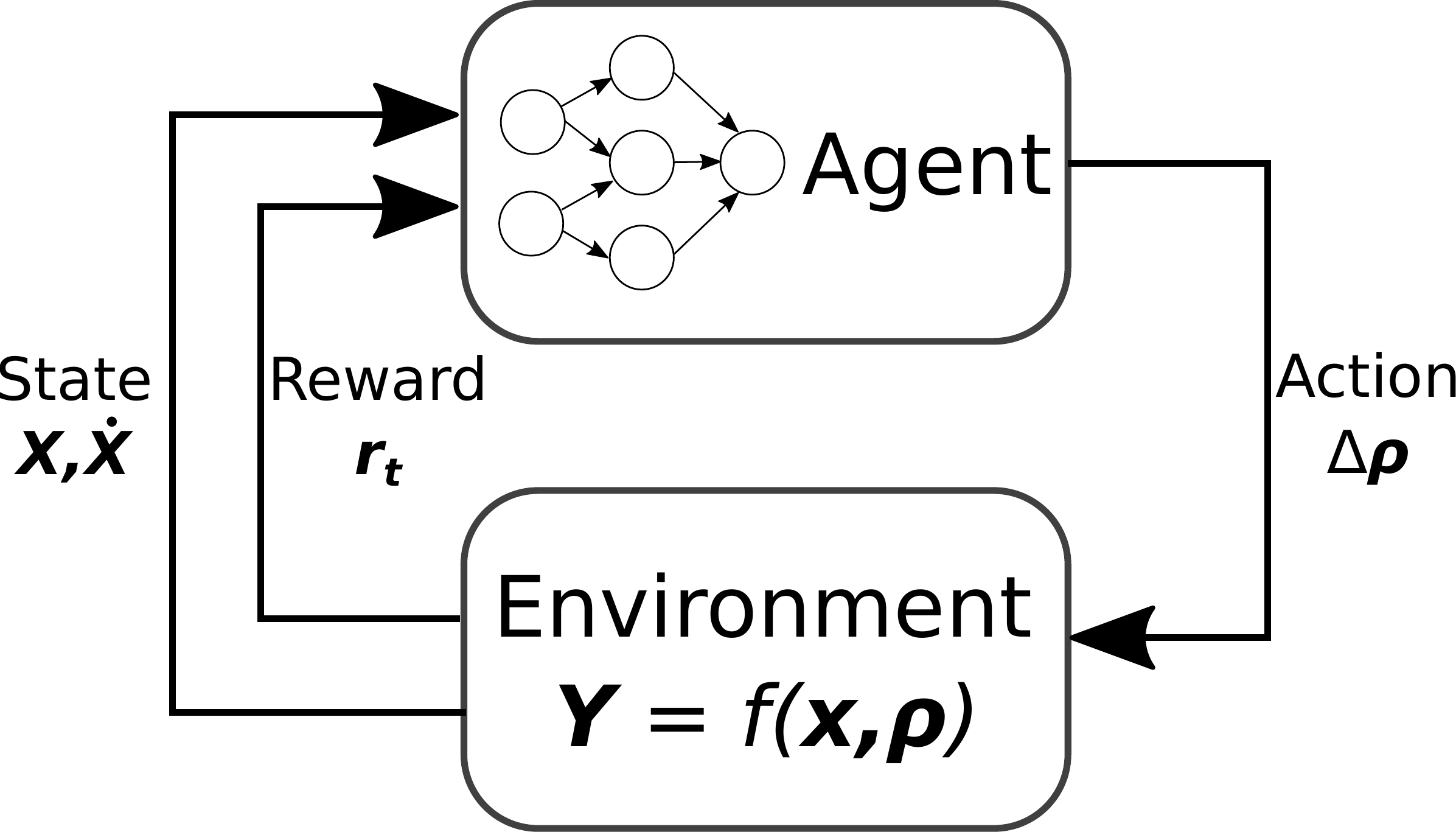}
\caption{Schematic illustrating the basic framework of a reinforcement learning problem. An agent continually perturbs the environment (the Lorenz equations in our case) by taking an action, and records the resulting states. The agent is rewarded when a desirable state is reached, and punished otherwise.}
\label{fig:rl_pic} 
\end{figure}

\begin{figure}
\subfloat[\label{sfig:lorenz_full}]{%
  \includegraphics[width=0.7\linewidth]{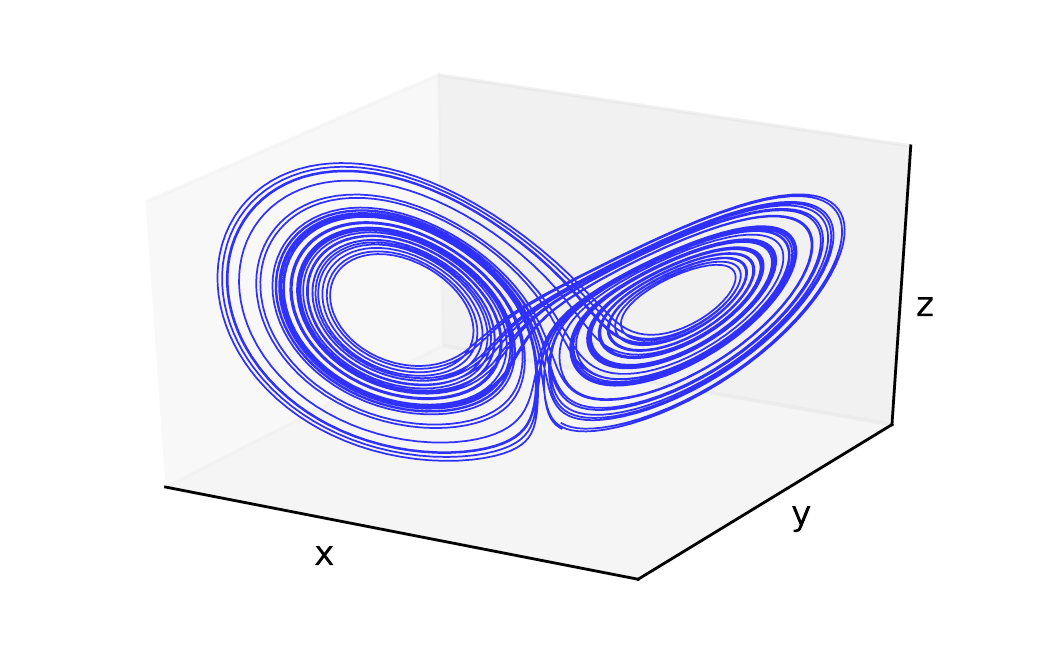}%
}\\
\subfloat[\label{sfig:lorenz_transient}]{%
  \includegraphics[width=0.7\linewidth]{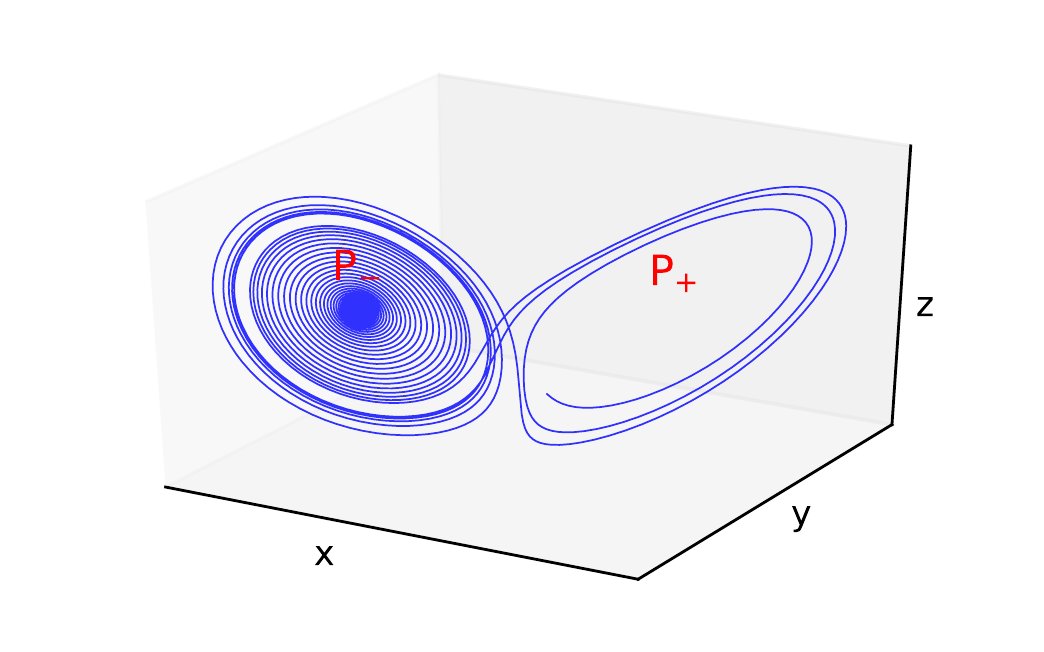}%
}
\caption{Solution of the Lorenz system of equations in (a) the chaotic regime with $\rho = 28$, and (b) the meta-stable chaotic regime with $\rho = 20$. Note that the solution traverses a chaotic trajectory in the first case, whereas it converges to $\mathrm{P_{-}}$ after a few chaotic transients in the second case.}
\label{fig:no_control} 
\end{figure}

As depicted in Fig.~\ref{fig:rl_pic}, a reinforcement learning problem consists of five major elements - a learning agent, an environment described by a model Y (the Lorenz system in our case), state-space S, action-space A, and reward {$r_t$}. Initially, the RL agent interacts with its environment in a trial-and-error manner. At each time step $t$, the agent receives the current state $s_t$ of the environment, and selects an action $a_t$ following a policy $\Pi(a_t|s_t)$. This action allows the agent to perturb the state of the environment, and move to a new state $s_{t+1}$ by evaluating the given model Y of the environment. Upon affecting this transition, the agent is rewarded (or punished) with reward $r_t$. This process continues until the agent reaches a terminal state, at which point a new episode starts over. The return received from each episode is the discounted cumlative reward with discount factor $\gamma$, which lies between $0$ and $1$. The discount factor makes it feasible to emphasize the importance of maximizing long-term reward, which enables the agent to prefer actions that are beneficial in the long-term. The cumulative reward, $R(a_t|s_t)$, is given as,
\begin{IEEEeqnarray}{rCl}
R(a_t|s_t) = \sum_{k=0}^{\infty} \gamma^{k}r_{t+k}
\end{IEEEeqnarray}
The goal of the RL agent is to maximize this cumulative reward by discovering an optimal policy $\Pi^{*}$.  There are a variety of methods available for attaining this. We make use of Proximal Policy Optimization (PPO)~\cite{schulman2017proximal} which is a type of policy Gradient Method (PGM) ~\cite{sutton2018reinforcement}. PPO is suitable for continuous-control problems~\cite{mnih2015human}, and it is simpler in its mathematical implementation compared to other PGM based RL algorithms~\cite{schulman2015trust}. Moreover, PPO requires comparitively little hyper-parameter tuning for use in a variety of different problems. The specific implementation of the algorithm that we used, PPO2, is available as part of the OpenAI stable-baselines library~\cite{stable-baselines}. The ergodic and unsteady nature of chaotic dynamics necessitates the use of a version of PPO2 wherein the policy is defined by deep recurrent neural networks comprised of long-short-term memory cells, instead of traditional feed-forward neural networks.
The environment for the Lorenz system is written in a OpenAI gym~\cite{brockman2016openai} -compatible python format, and is provided as part of the supplementary materials. The relevant equations are given as,
\begin{IEEEeqnarray}{rCl}
\label{eq:Lorenz}
\frac{dx}{dt}&=&\sigma(y-x) \IEEEyesnumber \IEEEyessubnumber* \label{subeq:Lorenz_a} \\
\frac{dy}{dt}&=&x(\rho - z) - y \label{subeq:Lorenz_b} \\
\frac{dz}{dt}&=&xy - \beta z \label{subeq:Lorenz_c}
\end{IEEEeqnarray}
With $\sigma=10$ and $\beta=8/3$, $\rho=28$ gives rise to chaotic trajectories, whereas transient chaos is found in the interval $\rho \in [13.93, 24.06]$ \cite{kaplan1979preturbulence}. Without any control implemented, the solution will converge to specific fix-points after a short transient, as shown in Fig.~\ref{fig:no_control}. The two fix-points in our case are given by $\mathrm{P_{+}}$ = $(7.12, 7.12, 19)$ and $\mathrm{P_{-}}$ = $(-7.12, -7.12, 19)$.

We use reinforcement learning to prevent such a transient from chaotic- to fix-point-solutions. This is done by perturbing the parameters in Eqs.~\ref{eq:Lorenz} ($\bm{\rho} = (\sigma, \rho, \beta)$) by ${\Delta\bm{\rho}}= (\Delta\sigma, \Delta\rho, \Delta\beta)$, with ${\Delta\bm{\rho}}\in[\bm{-\rho}/10, \bm{\rho}/10]$. The instantaneous value of the solution vector ${\bm{X}(t)} = (x,y,z)$ and its time-derivative (velocity) $ {\bm{\dot X}(t)} = ({{V_x}(t)}, {{V_y}(t)}, {{V_z}(t)}) = (\frac{dx}{dt}, \frac{dy}{dt}, \frac{dz}{dt})$ constitute the state space S for the RL algorithm. For training the RL agent to retain a chaotic trajectory, we utilize the fact that $|{\bm{V}(t)}|$ will decrease consistently as the solution converges to one of the fix-points, eventually becoming zero. On the other hand, $|{\bm{V}(t)}|$ will have a non-zero average value when the solution traces the chaotic attractor. Thus, whenever the agent determines suitable action values ${\Delta\bm{\rho}}$ for which $|{\bm{V}(t)}|$ is maintained above the predefined threshold value ${V_{0}}=40$, it is rewarded, otherwise it is punished. In doing so over several iterations, the agent eventually learns to keep the trajectory chaotic. 

The reward allocated to the agent consists of two parts: a stepwise reward ${r_{t}}$ provided at each time step, and a one-time terminal reward ${r_{terminal}}$ given at the end of each episode. The two terms take the following form,
\begin{IEEEeqnarray}{rCl}
{ r_{t}} = 
\begin{cases}
10 & {V(t) > V_{0}}\\
-10 &  {V(t) \leqslant V_{0}}\\
\end{cases} \IEEEyesnumber \IEEEyessubnumber*\\
{r_{terminal}} = 
\begin{cases}
-100 &{\bar{r}_{t} < -2 }\\
0 & {\bar{r}_{t} > -2 }
\end{cases}
\end{IEEEeqnarray}
The average ${\bar{r}_{t}}$ is defined over the last 2000 time steps of an episode, and facilitates learning to keep the trajectory chaotic over long periods of time. The training of the agent is divided into episodes of 4000 time steps each, with time step size $dt = 2e{-}2$. The RL agent is expected to learn suitable action values ${\Delta\bm{\rho}}$ for any state permissible by the system environment, such that the long-term reward accumulated is maximized.
\begin{figure}[!ht]
\includegraphics[scale=0.6]{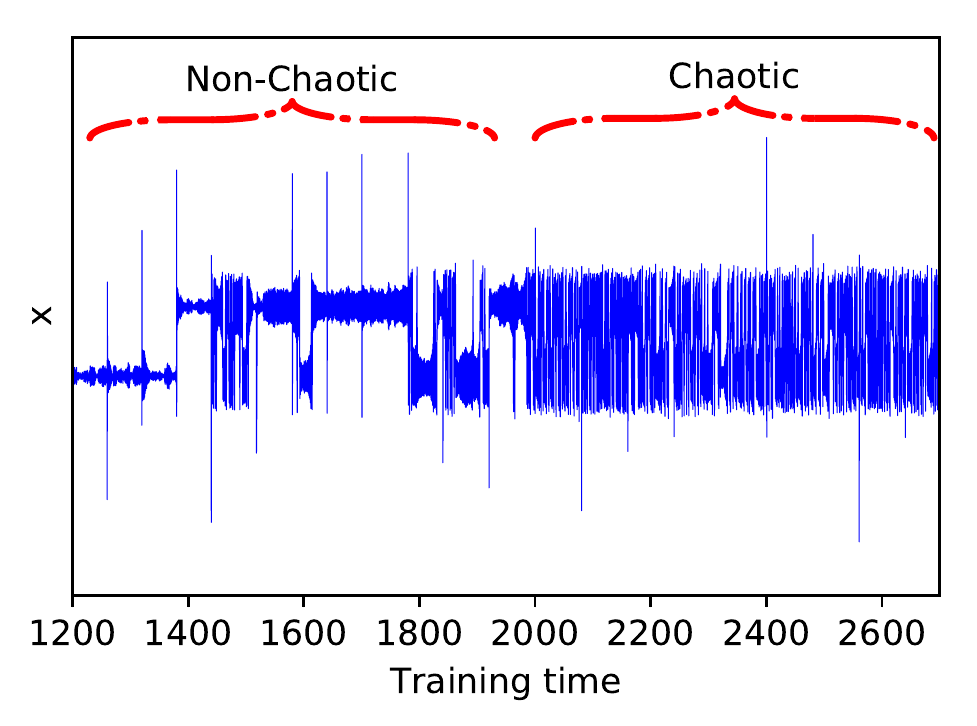}
\caption{Training of the RL agent with time. Note that the solution is non-chaotic until around t = 2000, beyond which the agent is able to take effective decisions to keep the solution-trajectory chaotic for further instances.}
\label{fig:training} 
\end{figure}
Fig.~\ref{fig:training} illustrates the training of the RL agent with time. The underlying neural network is trained for $2\times10^{5}$  time steps, which corresponds to 50 independent episodes in total, with each episode beginning with random values of the state variables $\bm{X}$ between -40 and 40; the corresponding values for $\bm{\dot X}$ are determined using the Lorenz equations (Eqs.~\ref{eq:Lorenz}). Initially, the solution keeps converging to the fix-points, since the network is unable to provide optimal action-decisions. After the network has trained for some time, it successfully learns the optimal actions for keeping the value of $|{\bm{V}(t)}|$ above ${V_{0}}$. As a consequence, the agent learns that the best way of maximizing reward is by maintaining the dynamics over the chaotic-attractor, which, although non-attracting for the given set of parameters, is a natural solution of the system.
\begin{figure}[ht]
\subfloat[\label{sfig:perturb}]{%
	\includegraphics[width=0.8\linewidth]{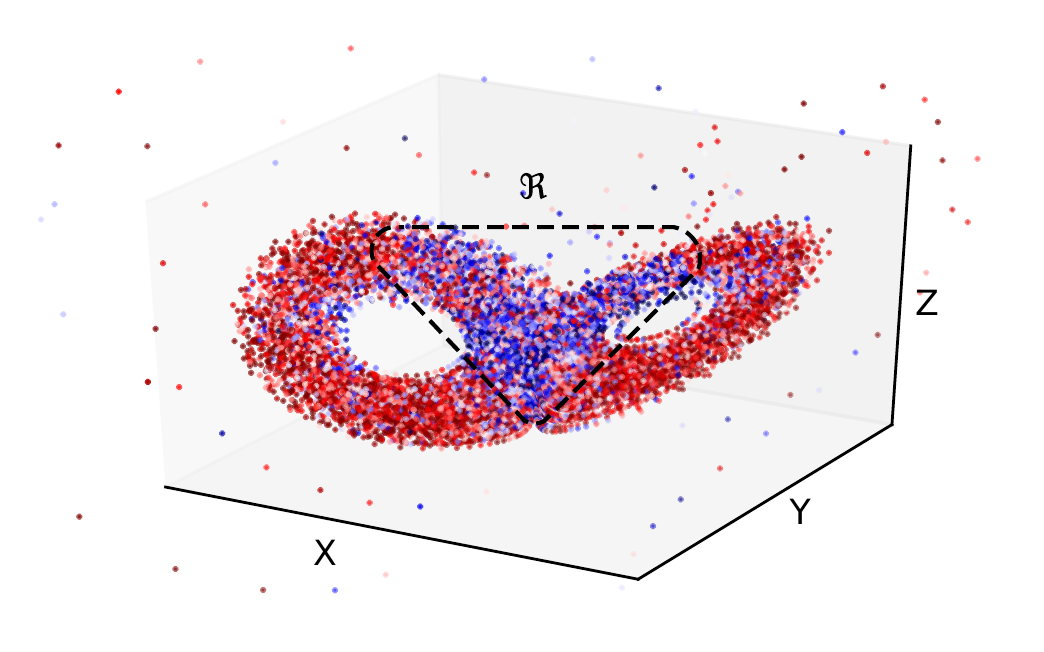}
}\\
\subfloat[\label{sfig:velocity}]{%
	\includegraphics[width=0.6\linewidth]{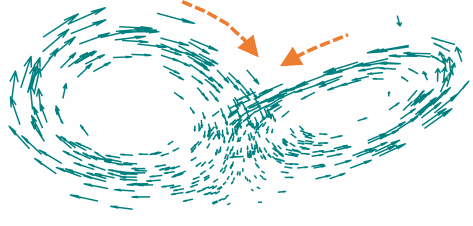}%
}
\caption{(a) Distribution of the perturbation $\Delta\rho$ learned by the RL agent to keep the dynamics on the chaotic-attractor. The red dots indicate locations where the perturbation values are positive, and the blue dots correspond to negative values. (b) Velocity vectors for the corresponding solution in the state space. Note that $\Delta\rho$ is predominantly negative in the region $\Re$ where $ {V_{z}} <0$.}
\label{fig:pert_rho} 
\end{figure}

Figure~\ref{fig:pert_rho} shows the distribution of the perturbations $\Delta\rho$ employed by the trained agent, which allow it to keep the dynamics on the chaotic-attractor. This distribution was obtained by plotting the controlled-trajectories for 400 random initial values for the variables $x$, $y$ and $z$, lying between -40 and 40. Note that a similar distribution was obtained for the other perturbations $\Delta\beta$ and $\Delta\sigma$. However, we find that an execution of the converged RL control-policy with $\Delta\beta$ and $\Delta\sigma$ explicitly set to zero does not make a difference in the control outcome; the agent is still able to maintain a chaotic trajectory. This may be attributed to the dominating magnitude of the parameter $\rho$ compared to the other two parameters.

As depicted in Fig.~\ref{fig:pert_rho}(a,b), the perturbation values are predominantly negative in the region $\Re$ where ${V_z}<0$ and positive elsewhere.  The success of this control-policy $\Pi^{*}$ in keeping the trajectory over the chaotic-attractor can be explained using the sensitivity of the solutions of Eqs.\ref{eq:Lorenz} to the perturbation $\Delta\rho$. For $x<0$ in $\Re$, a negative $\Delta\rho$ makes ${V_y}$ in Eq.(\ref{subeq:Lorenz_b}) more positive, which in turn makes y and hence ${V_x}$ in Eq.(\ref{subeq:Lorenz_a}) more positive, thus drifting the trajectory away from the fix-point $\mathrm{P_{-}}$. Similarly, when $x>0$ in $\Re$, a negative $\Delta\rho$ makes ${V_y}$ in Eq.(\ref{subeq:Lorenz_b}) more negative, which in turn makes y and hence ${V_x}$ more negative, thereby drifting the trajectory away from the fix-point $\mathrm{P_{+}}$. The role of positive perturbations outside $\Re$ in avoiding the escape of the trajectory from the chaotic attractor to the fix-points can be explained likewise. Positive perturbations of $\rho$ lead to an increase in $V_y$ (Eq.~\ref{subeq:Lorenz_b}). The subsequent increase in $y$ leads to an increase in $V_x$ (Eq.~\ref{subeq:Lorenz_a}), and the higher values of $x$ and $y$ lead to an increase in $V_z$ (Eq.~\ref{subeq:Lorenz_c}). The overall effect is to increase the speed, which prevents the trajectory from spiralling in to the fix-points.
\begin{figure}[!ht]
\includegraphics[width=0.7\linewidth]{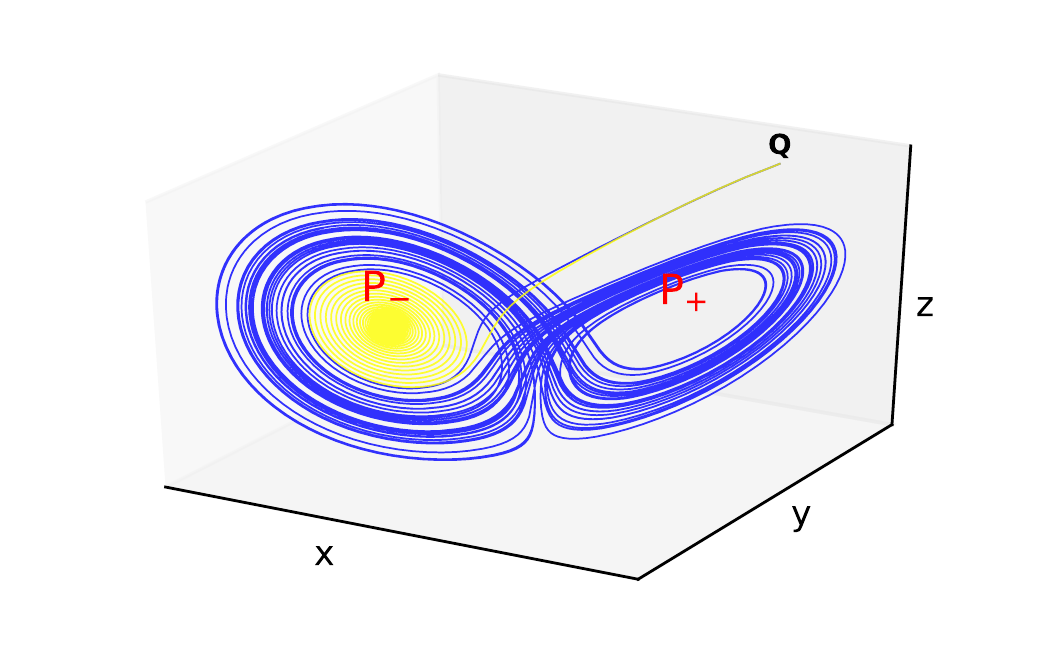}
\caption{Comparison of the trajectory with and without the application of rule-based control. The blue trajectory corresponds to the controlled solution, starting from the initial condition $\mathbf{Q}=(10,15,35)$. The yellow uncontrolled solution starts from the same initial condition, and spirals in to the fix-point $\mathrm{P_{-}}$.}
\label{fig:simple_control} 
\end{figure}

Based on this strategy, we formulate a simple rule-based controller which perturbs the parameter $\rho$ by $-\rho/10$ whenever the trajectory visits the region $\Re$, i.e., whenever $V_z<0$. All parameters remain unperturbed outside this region. The success of the rule-based binary-control is demonstrated in Fig.~\ref{fig:simple_control}, where the uncontrolled trajectory (yellow) converges to the fix-point, whereas its controlled counterpart (blue) remains chaotic. Note that, unlike other control-techniques, RL-based control requires no a-priori analytical knowledge about the dynamical system regarding its escape-regions, target-points and  safe-sets. The RL agent learns the optimal strategy {$\Pi^{*}$} to prevent the transition from chaotic to fix-point solutions completely autonomously, by continually interacting with the environment defined by the Lorenz system of equations exhibiting transient-chaos.

To conclude, we have demonstrated the utility of deep reinforcement learning in restoring chaos for a transiently-chaotic system. Since, transient chaos is a consequence of a catastrophic bifurcation (crisis)~\cite{wang2011predicting}, our results pave the way for RL enabled control of extreme-events and catastrophes in non-linear dynamical systems.

\bibliography{chaos_bibliography.bib}

\begin{thebibliography}{24}%
\makeatletter
\providecommand \@ifxundefined [1]{%
 \@ifx{#1\undefined}
}%
\providecommand \@ifnum [1]{%
 \ifnum #1\expandafter \@firstoftwo
 \else \expandafter \@secondoftwo
 \fi
}%
\providecommand \@ifx [1]{%
 \ifx #1\expandafter \@firstoftwo
 \else \expandafter \@secondoftwo
 \fi
}%
\providecommand \natexlab [1]{#1}%
\providecommand \enquote  [1]{``#1''}%
\providecommand \bibnamefont  [1]{#1}%
\providecommand \bibfnamefont [1]{#1}%
\providecommand \citenamefont [1]{#1}%
\providecommand \href@noop [0]{\@secondoftwo}%
\providecommand \href [0]{\begingroup \@sanitize@url \@href}%
\providecommand \@href[1]{\@@startlink{#1}\@@href}%
\providecommand \@@href[1]{\endgroup#1\@@endlink}%
\providecommand \@sanitize@url [0]{\catcode `\\12\catcode `\$12\catcode
  `\&12\catcode `\#12\catcode `\^12\catcode `\_12\catcode `\%12\relax}%
\providecommand \@@startlink[1]{}%
\providecommand \@@endlink[0]{}%
\providecommand \url  [0]{\begingroup\@sanitize@url \@url }%
\providecommand \@url [1]{\endgroup\@href {#1}{\urlprefix }}%
\providecommand \urlprefix  [0]{URL }%
\providecommand \Eprint [0]{\href }%
\providecommand \doibase [0]{http://dx.doi.org/}%
\providecommand \selectlanguage [0]{\@gobble}%
\providecommand \bibinfo  [0]{\@secondoftwo}%
\providecommand \bibfield  [0]{\@secondoftwo}%
\providecommand \translation [1]{[#1]}%
\providecommand \BibitemOpen [0]{}%
\providecommand \bibitemStop [0]{}%
\providecommand \bibitemNoStop [0]{.\EOS\space}%
\providecommand \EOS [0]{\spacefactor3000\relax}%
\providecommand \BibitemShut  [1]{\csname bibitem#1\endcsname}%
\let\auto@bib@innerbib\@empty
%</preamble>
\bibitem [{\citenamefont {Georgiou}\ and\ \citenamefont
  {Schwartz}(1996)}]{georgiou1996slow}%
  \BibitemOpen
  \bibfield  {author} {\bibinfo {author} {\bibfnamefont {I.~T.}\ \bibnamefont
  {Georgiou}}\ and\ \bibinfo {author} {\bibfnamefont {I.~B.}\ \bibnamefont
  {Schwartz}},\ }\href@noop {} {\bibfield  {journal} {\bibinfo  {journal} {Int.
  J. Bifurcation Chaos}\ }\textbf {\bibinfo {volume} {6}},\ \bibinfo {pages}
  {673} (\bibinfo {year} {1996})}\BibitemShut {NoStop}%
\bibitem [{\citenamefont {Haken}(1981)}]{haken1981chaos}%
  \BibitemOpen
  \bibfield  {author} {\bibinfo {author} {\bibfnamefont {H.}~\bibnamefont
  {Haken}},\ }in\ \href@noop {} {\emph {\bibinfo {booktitle} {Chaos and order
  in nature}}}\ (\bibinfo  {publisher} {Springer},\ \bibinfo {year} {1981})\
  pp.\ \bibinfo {pages} {2--11}\BibitemShut {NoStop}%
\bibitem [{\citenamefont {Sackellares}\ \emph {et~al.}(2000)\citenamefont
  {Sackellares}, \citenamefont {Iasemidis}, \citenamefont {Shiau},
  \citenamefont {Gilmore},\ and\ \citenamefont
  {Roper}}]{sackellares2000epilepsy}%
  \BibitemOpen
  \bibfield  {author} {\bibinfo {author} {\bibfnamefont {J.~C.}\ \bibnamefont
  {Sackellares}}, \bibinfo {author} {\bibfnamefont {L.~D.}\ \bibnamefont
  {Iasemidis}}, \bibinfo {author} {\bibfnamefont {D.-S.}\ \bibnamefont
  {Shiau}}, \bibinfo {author} {\bibfnamefont {R.~L.}\ \bibnamefont {Gilmore}},
  \ and\ \bibinfo {author} {\bibfnamefont {S.~N.}\ \bibnamefont {Roper}},\ }in\
  \href@noop {} {\emph {\bibinfo {booktitle} {Chaos in Brain?}}}\ (\bibinfo
  {publisher} {World Scientific},\ \bibinfo {year} {2000})\ pp.\ \bibinfo
  {pages} {112--133}\BibitemShut {NoStop}%
\bibitem [{\citenamefont {Goldberger}\ \emph {et~al.}(1990)\citenamefont
  {Goldberger}, \citenamefont {Rigney},\ and\ \citenamefont
  {West}}]{goldberger1990chaos}%
  \BibitemOpen
  \bibfield  {author} {\bibinfo {author} {\bibfnamefont {A.~L.}\ \bibnamefont
  {Goldberger}}, \bibinfo {author} {\bibfnamefont {D.~R.}\ \bibnamefont
  {Rigney}}, \ and\ \bibinfo {author} {\bibfnamefont {B.~J.}\ \bibnamefont
  {West}},\ }\href@noop {} {\bibfield  {journal} {\bibinfo  {journal} {Sci.
  Am.}\ }\textbf {\bibinfo {volume} {262}},\ \bibinfo {pages} {42} (\bibinfo
  {year} {1990})}\BibitemShut {NoStop}%
\bibitem [{\citenamefont {Yang}\ \emph {et~al.}(1995)\citenamefont {Yang},
  \citenamefont {Ding}, \citenamefont {Mandell},\ and\ \citenamefont
  {Ott}}]{yang1995preserving}%
  \BibitemOpen
  \bibfield  {author} {\bibinfo {author} {\bibfnamefont {W.}~\bibnamefont
  {Yang}}, \bibinfo {author} {\bibfnamefont {M.}~\bibnamefont {Ding}}, \bibinfo
  {author} {\bibfnamefont {A.~J.}\ \bibnamefont {Mandell}}, \ and\ \bibinfo
  {author} {\bibfnamefont {E.}~\bibnamefont {Ott}},\ }\href@noop {} {\bibfield
  {journal} {\bibinfo  {journal} {Phys. Rev. E}\ }\textbf {\bibinfo {volume}
  {51}},\ \bibinfo {pages} {102} (\bibinfo {year} {1995})}\BibitemShut
  {NoStop}%
\bibitem [{\citenamefont {Grebogi}\ \emph {et~al.}(1983)\citenamefont
  {Grebogi}, \citenamefont {Ott},\ and\ \citenamefont
  {Yorke}}]{grebogi1983crises}%
  \BibitemOpen
  \bibfield  {author} {\bibinfo {author} {\bibfnamefont {C.}~\bibnamefont
  {Grebogi}}, \bibinfo {author} {\bibfnamefont {E.}~\bibnamefont {Ott}}, \ and\
  \bibinfo {author} {\bibfnamefont {J.~A.}\ \bibnamefont {Yorke}},\ }\href@noop
  {} {\bibfield  {journal} {\bibinfo  {journal} {Physica D: Nonlinear
  Phenomena}\ }\textbf {\bibinfo {volume} {7}},\ \bibinfo {pages} {181}
  (\bibinfo {year} {1983})}\BibitemShut {NoStop}%
\bibitem [{\citenamefont {Dobson}\ and\ \citenamefont
  {Chiang}(1989)}]{dobson1989towards}%
  \BibitemOpen
  \bibfield  {author} {\bibinfo {author} {\bibfnamefont {I.}~\bibnamefont
  {Dobson}}\ and\ \bibinfo {author} {\bibfnamefont {H.-D.}\ \bibnamefont
  {Chiang}},\ }\href@noop {} {\bibfield  {journal} {\bibinfo  {journal}
  {Systems \& Control Lett.}\ }\textbf {\bibinfo {volume} {13}},\ \bibinfo
  {pages} {253} (\bibinfo {year} {1989})}\BibitemShut {NoStop}%
\bibitem [{\citenamefont {McCann}\ and\ \citenamefont
  {Yodzis}(1994)}]{mccann1994nonlinear}%
  \BibitemOpen
  \bibfield  {author} {\bibinfo {author} {\bibfnamefont {K.}~\bibnamefont
  {McCann}}\ and\ \bibinfo {author} {\bibfnamefont {P.}~\bibnamefont
  {Yodzis}},\ }\href@noop {} {\bibfield  {journal} {\bibinfo  {journal} {The
  American Naturalist}\ }\textbf {\bibinfo {volume} {144}},\ \bibinfo {pages}
  {873} (\bibinfo {year} {1994})}\BibitemShut {NoStop}%
\bibitem [{\citenamefont {Avila}\ \emph {et~al.}(2011)\citenamefont {Avila},
  \citenamefont {Moxey}, \citenamefont {de~Lozar}, \citenamefont {Avila},
  \citenamefont {Barkley},\ and\ \citenamefont {Hof}}]{avila2011onset}%
  \BibitemOpen
  \bibfield  {author} {\bibinfo {author} {\bibfnamefont {K.}~\bibnamefont
  {Avila}}, \bibinfo {author} {\bibfnamefont {D.}~\bibnamefont {Moxey}},
  \bibinfo {author} {\bibfnamefont {A.}~\bibnamefont {de~Lozar}}, \bibinfo
  {author} {\bibfnamefont {M.}~\bibnamefont {Avila}}, \bibinfo {author}
  {\bibfnamefont {D.}~\bibnamefont {Barkley}}, \ and\ \bibinfo {author}
  {\bibfnamefont {B.}~\bibnamefont {Hof}},\ }\href@noop {} {\bibfield
  {journal} {\bibinfo  {journal} {Science}\ }\textbf {\bibinfo {volume}
  {333}},\ \bibinfo {pages} {192} (\bibinfo {year} {2011})}\BibitemShut
  {NoStop}%
\bibitem [{\citenamefont {Kreilos}\ \emph {et~al.}(2014)\citenamefont
  {Kreilos}, \citenamefont {Eckhardt},\ and\ \citenamefont
  {Schneider}}]{kreilos2014increasing}%
  \BibitemOpen
  \bibfield  {author} {\bibinfo {author} {\bibfnamefont {T.}~\bibnamefont
  {Kreilos}}, \bibinfo {author} {\bibfnamefont {B.}~\bibnamefont {Eckhardt}}, \
  and\ \bibinfo {author} {\bibfnamefont {T.~M.}\ \bibnamefont {Schneider}},\
  }\href@noop {} {\bibfield  {journal} {\bibinfo  {journal} {Phys. Rev. Lett.}\
  }\textbf {\bibinfo {volume} {112}},\ \bibinfo {pages} {044503} (\bibinfo
  {year} {2014})}\BibitemShut {NoStop}%
\bibitem [{\citenamefont {Schwartz}\ and\ \citenamefont
  {Triandaf}(1996)}]{schwartz1996sustaining}%
  \BibitemOpen
  \bibfield  {author} {\bibinfo {author} {\bibfnamefont {I.~B.}\ \bibnamefont
  {Schwartz}}\ and\ \bibinfo {author} {\bibfnamefont {I.}~\bibnamefont
  {Triandaf}},\ }\href@noop {} {\bibfield  {journal} {\bibinfo  {journal}
  {Phys. Rev. Lett.}\ }\textbf {\bibinfo {volume} {77}},\ \bibinfo {pages}
  {4740} (\bibinfo {year} {1996})}\BibitemShut {NoStop}%
\bibitem [{\citenamefont {Dhamala}\ and\ \citenamefont
  {Lai}(1999)}]{dhamala1999controlling}%
  \BibitemOpen
  \bibfield  {author} {\bibinfo {author} {\bibfnamefont {M.}~\bibnamefont
  {Dhamala}}\ and\ \bibinfo {author} {\bibfnamefont {Y.-C.}\ \bibnamefont
  {Lai}},\ }\href@noop {} {\bibfield  {journal} {\bibinfo  {journal} {Phys.
  Rev. E}\ }\textbf {\bibinfo {volume} {59}},\ \bibinfo {pages} {1646}
  (\bibinfo {year} {1999})}\BibitemShut {NoStop}%
\bibitem [{\citenamefont {Cape{\'a}ns}\ \emph {et~al.}(2017)\citenamefont
  {Cape{\'a}ns}, \citenamefont {Sabuco}, \citenamefont {Sanju{\'a}n},\ and\
  \citenamefont {Yorke}}]{capeans2017partially}%
  \BibitemOpen
  \bibfield  {author} {\bibinfo {author} {\bibfnamefont {R.}~\bibnamefont
  {Cape{\'a}ns}}, \bibinfo {author} {\bibfnamefont {J.}~\bibnamefont {Sabuco}},
  \bibinfo {author} {\bibfnamefont {M.~A.}\ \bibnamefont {Sanju{\'a}n}}, \ and\
  \bibinfo {author} {\bibfnamefont {J.~A.}\ \bibnamefont {Yorke}},\ }\href@noop
  {} {\bibfield  {journal} {\bibinfo  {journal} {Philos. Trans. R. Soc. London,
  Ser. A}\ }\textbf {\bibinfo {volume} {375}},\ \bibinfo {pages} {20160211}
  (\bibinfo {year} {2017})}\BibitemShut {NoStop}%
\bibitem [{\citenamefont {Silver}\ \emph {et~al.}(2016)\citenamefont {Silver},
  \citenamefont {Huang}, \citenamefont {Maddison}, \citenamefont {Guez},
  \citenamefont {Sifre}, \citenamefont {Van Den~Driessche}, \citenamefont
  {Schrittwieser}, \citenamefont {Antonoglou}, \citenamefont {Panneershelvam},
  \citenamefont {Lanctot} \emph {et~al.}}]{silver2016mastering}%
  \BibitemOpen
  \bibfield  {author} {\bibinfo {author} {\bibfnamefont {D.}~\bibnamefont
  {Silver}}, \bibinfo {author} {\bibfnamefont {A.}~\bibnamefont {Huang}},
  \bibinfo {author} {\bibfnamefont {C.~J.}\ \bibnamefont {Maddison}}, \bibinfo
  {author} {\bibfnamefont {A.}~\bibnamefont {Guez}}, \bibinfo {author}
  {\bibfnamefont {L.}~\bibnamefont {Sifre}}, \bibinfo {author} {\bibfnamefont
  {G.}~\bibnamefont {Van Den~Driessche}}, \bibinfo {author} {\bibfnamefont
  {J.}~\bibnamefont {Schrittwieser}}, \bibinfo {author} {\bibfnamefont
  {I.}~\bibnamefont {Antonoglou}}, \bibinfo {author} {\bibfnamefont
  {V.}~\bibnamefont {Panneershelvam}}, \bibinfo {author} {\bibfnamefont
  {M.}~\bibnamefont {Lanctot}},  \emph {et~al.},\ }\href@noop {} {\bibfield
  {journal} {\bibinfo  {journal} {Nature}\ }\textbf {\bibinfo {volume} {529}},\
  \bibinfo {pages} {484} (\bibinfo {year} {2016})}\BibitemShut {NoStop}%
\bibitem [{\citenamefont {Verma}\ \emph {et~al.}(2018)\citenamefont {Verma},
  \citenamefont {Novati},\ and\ \citenamefont
  {Koumoutsakos}}]{verma2018efficient}%
  \BibitemOpen
  \bibfield  {author} {\bibinfo {author} {\bibfnamefont {S.}~\bibnamefont
  {Verma}}, \bibinfo {author} {\bibfnamefont {G.}~\bibnamefont {Novati}}, \
  and\ \bibinfo {author} {\bibfnamefont {P.}~\bibnamefont {Koumoutsakos}},\
  }\href@noop {} {\bibfield  {journal} {\bibinfo  {journal} {Proc. Natl. Acad.
  Sci. USA}\ }\textbf {\bibinfo {volume} {115}},\ \bibinfo {pages} {5849}
  (\bibinfo {year} {2018})}\BibitemShut {NoStop}%
\bibitem [{\citenamefont {Lorenz}(1995)}]{lorenz1995essence}%
  \BibitemOpen
  \bibfield  {author} {\bibinfo {author} {\bibfnamefont {E.~N.}\ \bibnamefont
  {Lorenz}},\ }\href@noop {} {\emph {\bibinfo {title} {The essence of chaos}}}\
  (\bibinfo  {publisher} {University of Washington Press},\ \bibinfo {year}
  {1995})\BibitemShut {NoStop}%
\bibitem [{\citenamefont {Schulman}\ \emph {et~al.}(2017)\citenamefont
  {Schulman}, \citenamefont {Wolski}, \citenamefont {Dhariwal}, \citenamefont
  {Radford},\ and\ \citenamefont {Klimov}}]{schulman2017proximal}%
  \BibitemOpen
  \bibfield  {author} {\bibinfo {author} {\bibfnamefont {J.}~\bibnamefont
  {Schulman}}, \bibinfo {author} {\bibfnamefont {F.}~\bibnamefont {Wolski}},
  \bibinfo {author} {\bibfnamefont {P.}~\bibnamefont {Dhariwal}}, \bibinfo
  {author} {\bibfnamefont {A.}~\bibnamefont {Radford}}, \ and\ \bibinfo
  {author} {\bibfnamefont {O.}~\bibnamefont {Klimov}},\ }\href@noop {}
  {\bibfield  {journal} {\bibinfo  {journal} {arXiv preprint arXiv:1707.06347}\
  } (\bibinfo {year} {2017})}\BibitemShut {NoStop}%
\bibitem [{\citenamefont {Sutton}\ and\ \citenamefont
  {Barto}(2018)}]{sutton2018reinforcement}%
  \BibitemOpen
  \bibfield  {author} {\bibinfo {author} {\bibfnamefont {R.~S.}\ \bibnamefont
  {Sutton}}\ and\ \bibinfo {author} {\bibfnamefont {A.~G.}\ \bibnamefont
  {Barto}},\ }\href@noop {} {\emph {\bibinfo {title} {Reinforcement learning:
  An introduction}}}\ (\bibinfo  {publisher} {MIT press},\ \bibinfo {year}
  {2018})\BibitemShut {NoStop}%
\bibitem [{\citenamefont {Mnih}\ \emph {et~al.}(2015)\citenamefont {Mnih},
  \citenamefont {Kavukcuoglu}, \citenamefont {Silver}, \citenamefont {Rusu},
  \citenamefont {Veness}, \citenamefont {Bellemare}, \citenamefont {Graves},
  \citenamefont {Riedmiller}, \citenamefont {Fidjeland}, \citenamefont
  {Ostrovski} \emph {et~al.}}]{mnih2015human}%
  \BibitemOpen
  \bibfield  {author} {\bibinfo {author} {\bibfnamefont {V.}~\bibnamefont
  {Mnih}}, \bibinfo {author} {\bibfnamefont {K.}~\bibnamefont {Kavukcuoglu}},
  \bibinfo {author} {\bibfnamefont {D.}~\bibnamefont {Silver}}, \bibinfo
  {author} {\bibfnamefont {A.~A.}\ \bibnamefont {Rusu}}, \bibinfo {author}
  {\bibfnamefont {J.}~\bibnamefont {Veness}}, \bibinfo {author} {\bibfnamefont
  {M.~G.}\ \bibnamefont {Bellemare}}, \bibinfo {author} {\bibfnamefont
  {A.}~\bibnamefont {Graves}}, \bibinfo {author} {\bibfnamefont
  {M.}~\bibnamefont {Riedmiller}}, \bibinfo {author} {\bibfnamefont {A.~K.}\
  \bibnamefont {Fidjeland}}, \bibinfo {author} {\bibfnamefont {G.}~\bibnamefont
  {Ostrovski}},  \emph {et~al.},\ }\href@noop {} {\bibfield  {journal}
  {\bibinfo  {journal} {Nature}\ }\textbf {\bibinfo {volume} {518}},\ \bibinfo
  {pages} {529} (\bibinfo {year} {2015})}\BibitemShut {NoStop}%
\bibitem [{\citenamefont {Schulman}\ \emph {et~al.}(2015)\citenamefont
  {Schulman}, \citenamefont {Levine}, \citenamefont {Abbeel}, \citenamefont
  {Jordan},\ and\ \citenamefont {Moritz}}]{schulman2015trust}%
  \BibitemOpen
  \bibfield  {author} {\bibinfo {author} {\bibfnamefont {J.}~\bibnamefont
  {Schulman}}, \bibinfo {author} {\bibfnamefont {S.}~\bibnamefont {Levine}},
  \bibinfo {author} {\bibfnamefont {P.}~\bibnamefont {Abbeel}}, \bibinfo
  {author} {\bibfnamefont {M.}~\bibnamefont {Jordan}}, \ and\ \bibinfo {author}
  {\bibfnamefont {P.}~\bibnamefont {Moritz}},\ }in\ \href@noop {} {\emph
  {\bibinfo {booktitle} {ICML}}}\ (\bibinfo {year} {2015})\ pp.\ \bibinfo
  {pages} {1889--1897}\BibitemShut {NoStop}%
\bibitem [{\citenamefont {Hill}\ \emph {et~al.}(2018)\citenamefont {Hill},
  \citenamefont {Raffin}, \citenamefont {Ernestus}, \citenamefont {Gleave},
  \citenamefont {Traore}, \citenamefont {Dhariwal}, \citenamefont {Hesse},
  \citenamefont {Klimov}, \citenamefont {Nichol}, \citenamefont {Plappert},
  \citenamefont {Radford}, \citenamefont {Schulman}, \citenamefont {Sidor},\
  and\ \citenamefont {Wu}}]{stable-baselines}%
  \BibitemOpen
  \bibfield  {author} {\bibinfo {author} {\bibfnamefont {A.}~\bibnamefont
  {Hill}}, \bibinfo {author} {\bibfnamefont {A.}~\bibnamefont {Raffin}},
  \bibinfo {author} {\bibfnamefont {M.}~\bibnamefont {Ernestus}}, \bibinfo
  {author} {\bibfnamefont {A.}~\bibnamefont {Gleave}}, \bibinfo {author}
  {\bibfnamefont {R.}~\bibnamefont {Traore}}, \bibinfo {author} {\bibfnamefont
  {P.}~\bibnamefont {Dhariwal}}, \bibinfo {author} {\bibfnamefont
  {C.}~\bibnamefont {Hesse}}, \bibinfo {author} {\bibfnamefont
  {O.}~\bibnamefont {Klimov}}, \bibinfo {author} {\bibfnamefont
  {A.}~\bibnamefont {Nichol}}, \bibinfo {author} {\bibfnamefont
  {M.}~\bibnamefont {Plappert}}, \bibinfo {author} {\bibfnamefont
  {A.}~\bibnamefont {Radford}}, \bibinfo {author} {\bibfnamefont
  {J.}~\bibnamefont {Schulman}}, \bibinfo {author} {\bibfnamefont
  {S.}~\bibnamefont {Sidor}}, \ and\ \bibinfo {author} {\bibfnamefont
  {Y.}~\bibnamefont {Wu}},\ }\href@noop {} {\enquote {\bibinfo {title} {Stable
  baselines},}\ }\bibinfo {howpublished}
  {\url{https://github.com/hill-a/stable-baselines}} (\bibinfo {year}
  {2018})\BibitemShut {NoStop}%
\bibitem [{\citenamefont {Brockman}\ \emph {et~al.}(2016)\citenamefont
  {Brockman}, \citenamefont {Cheung}, \citenamefont {Pettersson}, \citenamefont
  {Schneider}, \citenamefont {Schulman}, \citenamefont {Tang},\ and\
  \citenamefont {Zaremba}}]{brockman2016openai}%
  \BibitemOpen
  \bibfield  {author} {\bibinfo {author} {\bibfnamefont {G.}~\bibnamefont
  {Brockman}}, \bibinfo {author} {\bibfnamefont {V.}~\bibnamefont {Cheung}},
  \bibinfo {author} {\bibfnamefont {L.}~\bibnamefont {Pettersson}}, \bibinfo
  {author} {\bibfnamefont {J.}~\bibnamefont {Schneider}}, \bibinfo {author}
  {\bibfnamefont {J.}~\bibnamefont {Schulman}}, \bibinfo {author}
  {\bibfnamefont {J.}~\bibnamefont {Tang}}, \ and\ \bibinfo {author}
  {\bibfnamefont {W.}~\bibnamefont {Zaremba}},\ }\href@noop {} {\bibfield
  {journal} {\bibinfo  {journal} {arXiv preprint arXiv:1606.01540}\ } (\bibinfo
  {year} {2016})}\BibitemShut {NoStop}%
\bibitem [{\citenamefont {Kaplan}\ and\ \citenamefont
  {Yorke}(1979)}]{kaplan1979preturbulence}%
  \BibitemOpen
  \bibfield  {author} {\bibinfo {author} {\bibfnamefont {J.~L.}\ \bibnamefont
  {Kaplan}}\ and\ \bibinfo {author} {\bibfnamefont {J.~A.}\ \bibnamefont
  {Yorke}},\ }\href@noop {} {\bibfield  {journal} {\bibinfo  {journal} {Comm.
  Math. Phys.}\ }\textbf {\bibinfo {volume} {67}},\ \bibinfo {pages} {93}
  (\bibinfo {year} {1979})}\BibitemShut {NoStop}%
\bibitem [{\citenamefont {Wang}\ \emph {et~al.}(2011)\citenamefont {Wang},
  \citenamefont {Yang}, \citenamefont {Lai}, \citenamefont {Kovanis},\ and\
  \citenamefont {Grebogi}}]{wang2011predicting}%
  \BibitemOpen
  \bibfield  {author} {\bibinfo {author} {\bibfnamefont {W.-X.}\ \bibnamefont
  {Wang}}, \bibinfo {author} {\bibfnamefont {R.}~\bibnamefont {Yang}}, \bibinfo
  {author} {\bibfnamefont {Y.-C.}\ \bibnamefont {Lai}}, \bibinfo {author}
  {\bibfnamefont {V.}~\bibnamefont {Kovanis}}, \ and\ \bibinfo {author}
  {\bibfnamefont {C.}~\bibnamefont {Grebogi}},\ }\href@noop {} {\bibfield
  {journal} {\bibinfo  {journal} {Phys. Rev. Lett.}\ }\textbf {\bibinfo
  {volume} {106}},\ \bibinfo {pages} {154101} (\bibinfo {year}
  {2011})}\BibitemShut {NoStop}%
\end{thebibliography}%

\end{document}